\documentclass[a4paper,11pt]{article}

\usepackage{amsmath}
\usepackage{graphicx}
\usepackage{amssymb}
\usepackage{a4wide}
\usepackage{authblk}
\usepackage{url}
\usepackage{setspace}

\newcommand{\unimoreaffil}{\small Department of Physics, Informatics and Mathematics, University of Modena and Reggio Emilia}
\newcommand{\uniboaffil}{\small Department of Computer Science and Engineering (DISI), University of Bologna}
\newcommand{\ecltaffil}{\small European Centre for Living Technology, Venezia, Italy}
\newcommand{\exploraaffil}{Explora s.r.l., Roma, Italy}
\title{Dynamical criticality: overview and open questions}

\author[1]{Andrea~Roli\thanks{Corresponding author, email: \texttt{andrea.roli@unibo.it}}}
\author[2,3]{Marco~Villani}
\author[4,3]{Alessandro~Filisetti}
\author[2,3]{Roberto~Serra}

\affil[1]{\uniboaffil}
\affil[2]{\unimoreaffil}
\affil[3]{\ecltaffil}
\affil[4]{\exploraaffil}

\date{}

\begin{document}

\maketitle

\doublespacing

\begin{abstract}
Systems that exhibit complex behaviours are often found in a particular dynamical condition, poised between order and disorder. This observation is at the core of the so-called \textit{criticality hypothesis}, which states that systems in a dynamical regime between order and disorder attain the highest level of computational capabilities and achieve an optimal trade-off between robustness and flexibility.
Recent results in cellular and evolutionary biology, neuroscience and computer science have revitalised the interest in the criticality hypothesis, emphasising its role as a viable candidate general law in adaptive complex systems.

In this paper we provide an overview of the works on dynamical criticality that are---to the best of our knowledge---particularly relevant for the criticality hypothesis. We review the main contributions concerning dynamics and information processing at the edge of chaos, and we illustrate the main achievements in the study of critical dynamics in biological systems. Finally, we discuss open questions and propose an agenda for future work.

\vspace{1ex}
\end{abstract}

%%%%%%%%%%%%%%%%%%%%%%%%%%%%%%%%%%%%%%%%%%%%%%%%%%%%%%%%%%%%%%%%%%%%%%%%%%%%%%%%%%%%%%%%%%%%%%%%%%%%%%%%%%%%%%%%%%%%%%%
\section{Introduction}

The peculiar properties of critical systems are at the roots of a conjecture stating that systems in a dynamical regime between order and disorder optimally balance robustness and adaptiveness, and reliably respond to inputs while being capable to react with a wide repertoire of possible actions. This conjecture was proposed by Kauffman~\cite{Kau1993,Kau1996} with main focus on living systems and by Packard, Langton and Crutchfield~\cite{Pac1988,Lan1990,CruYou1990} who introduced the expression ``computation at the edge of chaos''. In the last ten years, results have been presented providing evidence to support this conjecture in cellular biology and neuroscience. These results are not only encouraging, but they also suggest that the criticality hypothesis may play the role of a general law in adaptive complex systems dynamics. Nevertheless, recent works also enlighten some issues that have to be tackled so as to provide a solid formulation of the conjecture and fruitfully exploit it in both modelling biological systems and designing artificial ones. 

In this work, we provide a literature overview of this conjecture, selecting those works that are, to the best of our knowledge, particularly significant and we illustrate the main scientific questions addressed, along with open perspectives. 
It is important to observe that the term critical is used with slightly different meanings; in this review we will focus mainly on dynamical criticality, sometimes called ``the edge of chaos''.
This paper is organized as follows. In Section~\ref{sec:pt} we summarize preliminary notions on phase transitions and critical phenomena. Section~\ref{sec:edgeOfChaos} illustrates the main works concerning dynamics and computation at the edge of chaos. The criticality hypothesis is discussed in Section~\ref{sec:criticalLivingSystems} and in Section~\ref{sec:discussion} we discuss the state of the art and open questions. Section~\ref{sec:conclusion} concludes the work and outlook future work.

%%%%%%%%%%%%%%%%%%%%%%%%%%%%%%%%%%%%%%%%%%%%%%%%%%%%%%%%%%%%%%%%%%%%%%%%%%%%%%%%%%%%%%%%%%%%%%%%%%%%%%%%%%%%%%%%%%%%%%%
\section{Criticality and phase transitions}
\label{sec:pt}

Critical states have been first introduced in the theory of phase transitions, that describes phenomena in which a system undergoes a sharp change in some of its macroscopic properties if a suitable \emph{control parameter} is changed~\cite{BinDowFisNew1992,SolMan1996}. The macroscopic properties of the system are usually defined in terms of an \emph{order parameter}. For example, let us consider the familiar transition of $\textrm{H}_\textrm{2}\textrm{O}$ from water to gas: at constant pressure, the control parameter is the temperature $T$ and the order parameter $\Phi$ is the difference between the density of the substance and the density of gas. When $T < T_c = 100$ $^o$C, $\Phi$ is strictly positive, while for $T \ge T_c$ the parameter $\Phi$ is 0. Therefore, at $T = T_c$ the order parameter abruptly goes to zero. The point at which the transition occurs is said to be the \emph{critical point}~\cite{sole-PTbook}. The phase change requires some energy, called latent heat, which characterizes all the phase transitions of this kind, which are sometimes called \emph{first-order phase transitions} because it is the first derivative of the thermodynamic free energy (i.e. the order parameter) that is discontinuous at the transition point.
There are also extremely relevant  cases, where the order parameter changes continuously but some of its first-order derivatives change abruptly: these phase transitions are named \emph{second-order phase transitions} because the discontinuity affects the second derivatives of the free energy. However, it has been found that this distinction, which dates back to Paul Ehrenfest, has some limitations and it is often overlooked. A typical example of second-order phase transition is that of iron changing from paramagnetic to ferromagnetic state. When the temperature $T$ is greater than the so-called Curie temperature $T_c = 1043 \; K$, iron is paramagnetic. If an external magnetic field is imposed, then the overall magnetization $M$ of the material is proportional to the intensity of the external field. Conversely, for $T < T_c$ the material is magnetized even in the absence of an external field, to which it will tend to align if it is applied. In this case, heating iron from low to high temperature the magnetization simply goes to zero, without any sharp change. However, what changes discontinuously is the magnetization rate, which changes discontinuously at $T = T_c$.\\

The importance of phase transitions, especially the second-order ones, is due to some notable properties of the critical point.

The first property is \emph{universality}. It has been observed that order parameters can be described with power laws at the critical point. For example, in the proximity of the critical temperature $T_c$, the magnetization $M$ can be expressed as $M \sim (T - T_c)^{-\beta}$,  where $\beta>0$ is called the critical exponent. Surprisingly, the values of the critical exponents are indifferent to the details of the system and they have the same value for wide classes of systems, characterized by common topological and dimensional properties. Therefore, it is possible to classify systems subject to phase transitions in terms of universality classes, defined on the basis of the values of the critical exponents.

A second important property is that the influence between distant portions of the system is maximal at the critical point. More precisely, the average correlation length $\Gamma$, which measures the statistical correlation between any pair of elements in the system, follows a power law:  
$\Gamma \sim (T - T_c)^{-\xi}$,  with $\xi>0$. As a result, for finite systems the correlation between any pair of elements is maximal at the critical point.

Power laws assume a prominent role because they characterize the relevant quantities of the system at the critical point. For example, in a ferromagnetic model composed of atoms that can assume one out of two spins $(-1,+1)$, the distribution of clusters of homogeneous spins at the phase transition is described by a power law. This property does not hold if the system is away from the critical point. Moreover, it has been observed that the power spectrum of some key quantities decays as a power of the frequency, instead of showing the familiar exponential behaviour (the effect is sometimes called $1/f$ noise, although different powers of frequency may be involved).

Finally, the response of a system to external perturbations scales as a power law at the critical point; as a consequence, there is no characteristic scale of response to perturbations. Conversely, when the system is far from the critical point, the effect of perturbations can be expressed in terms of distributions with a characteristic scale. It is also worth to mention the phenomenon of \textit{critical slowing down}~\cite{Wis1984}, which consists in an asymptotically long time for a critical system to absorb the perturbation.

Phase transitions are usually studied by means of mean-field theory and renormalisation group theory~\cite{BinDowFisNew1992}. Recently, also techniques from information theory and information geometry have been successfully applied, as well as approaches that use Fisher information~\cite{JanJohKen2004,WanLizPro2011,ProLizObsWan2011,Pro2013}.\\

Phase transitions occur at precise values of the control parameters. Therefore, it is natural to ask the question as to why so many natural systems seem to settle exactly around the critical point, without a careful tuning of such parameters. One of the most successful attempts to answer this question comes from the models of self-organized criticality, SOC ~\cite{BakTanWie1987,Bak1996,Jen1998}. SOC systems tend spontaneously to ``self-organized critical states'', like in the case of the well-known sandpile model by Bak and co-workers. These states are called critical because they exhibit some of the characteristics observed at critical points, the most important one being a power-law behaviour of the fluctuations. The relations between SOC and the theory of phase transitions and criticality has been also investigated~\cite{SorJohDor1995,DicMuVesZap2000,BagPalRec1997,LuqBalMur2001}. SOC is certainly relevant for the study of complex systems, but in this review we are mainly concerned with the phenomenon of dynamical criticality: in this case, there are qualitatively different dynamical behaviours corresponding to different parameter values, and the critical points (or surfaces) separate regions in parameter space that correspond to different behaviours. This notion of criticality thus represents a straightforward generalization of the one that is used in describing phase transitions. The most interesting case is the one where there are regions of chaotic behaviours and regions of ordered (constant in time or regularly oscillating) behaviours. This case has also been called for quite obvious regions ``the edge of chaos'' and it will be the subject of the following sections.

It is important here to recall here the notion of ordered and disordered dynamical regimes of a system. A dynamical system in an ordered regime is characterized by stationary states that do not change in time, or that oscillate regularly. Moreover, when a system in the ordered regime is perturbed, the effect of such perturbation dies out. In a sense, systems in the ordered regime are robust and resilient against perturbations. Conversely, in a system in a disordered regime the effect of even a small perturbation spreads across the whole system. The steady states of a disordered system have no regular patterns and may appear as completely random. However, also deterministic systems can show disordered regimes, as in the case of chaotic dynamical systems, which have strange attractors~\cite{Str1994}. The hallmark of chaotic dynamics is an extreme sensitivity to initial conditions: slightly different initial conditions lead to exponentially fast divergent trajectories (even if both belonging to the same strange attractor). In the following, with a slight abuse of terminology, we will use the terms disordered and chaotic interchangeably. Critical systems are intermediate between these two cases and their steady states are characterized by a mixture of properties from ordered and disordered systems. In addition, when subject to external stimuli, the size of the perturbation remains constant on average for long time.
The change between order and chaos is also related to symmetry breaking and self-organization~\cite{Hak1977,NicPri1977,NicPri1989,Pol2007}.

%%%%%%%%%%%%%%%%%%%%%%%%%%%%%%%%%%%%%%%%%%%%%%%%%%%%%%%%%%%%%%%%%%%%%%%%%%%%%%%%%%%%%%%%%%%%%%%%%%%%%%%%%%%%%%%%%%%%%%%
\section{Computation at the edge of chaos}
\label{sec:edgeOfChaos}

The peculiar properties of critical systems enlightened in thermodynamics and statistical physics are at the roots of a conjecture stating that systems at the phase transition achieve the highest level of computational capability. The rationale behind this hypothesis is that ordered regimes are too rigid to be able to compute complex tasks, as changes are rapidly erased and the flow of information among the units of the system is rather low. Conversely, disordered regimes are too erratic to provide a reliable response to inputs, as perturbations and noise spread unboundedly, preventing effective information transmission and storage. Critical regimes may indeed provide the optimal trade-off between reliability and flexibility, i.e. they make the system able to react consistently with the inputs and, at the same time, capable to provide a sufficiently large number of possible outcomes.\\

This conjecture has been first proposed by Packard~\cite{Pac1988}, Langton~\cite{Lan1990} and Crutchfield~\cite{CruYou1990} who introduced the expression ``computation at the edge of chaos''. Langton studied the dynamical properties of cellular automata (CA). CA are systems composed of finite state automata, a.k.a. cells, arranged in a D-dimensional lattice. Each cell takes as inputs the states of the cells in a given neighbourhood. The transition function is supposed to be the same for all the cells. The simplest case is that of deterministic binary 1-dimensional CA, with neighbourhood composed of adjacent cells that update their state synchronously. Despite their apparent simplicity, these CA were shown to exhibit nontrivial behaviours, classified by Wolfram~\cite{Wol1984} into four classes: the first two classes are characterized by ``ordered'' CA, which evolve in time reaching a homogeneous state (Class~1) or a set of stable or periodic structures (Class~2). Class~3 is composed of CA showing a chaotic behaviour, i.e. sensitive to perturbations in the initial conditions. Finally, class~4 shows complex behaviours, exhibiting complex patterns in its time evolution. Langton defined a parameter, $\lambda$, that quantifies the equidistribution of states in the transition function: for $\lambda=0$, all transitions lead to one given state, whilst for $\lambda \approx 1$ all the possible transitions are equally represented in the transition function. Langton showed that $\lambda$ can play the role of a control parameter for CA and that the behaviour of CA moves from ordered to disordered as $\lambda$ increases approaching 1. The transition from order to disorder takes place for $\lambda \approx 0.5$ and is associated to properties of a second-order phase transition: critical slowing down, transients can be described with a power law and the average mutual information between cells is maximal. Remarkably, the critical value of $\lambda$ corresponds to transition functions belonging to class~4 according to Wolfram. It has also been shown that some CA in Wolfram class~4 are capable of universal computation, i.e. they are computationally equivalent to universal Turing machines (see, e.g.~\cite{Coo2004}). As a consequence, Langton conjectured that complex computational capabilities are attained at the phase transition.

Langton's conjecture finds a first principled theoretical support in the work by Crutchfiled (see, e.g.~\cite{CruYou1990}) in which the intrinsic computational properties of a system are estimated on the basis of stochastic automata. Given the system under observation, a data stream of its time evolution is used to build an $\epsilon$-machine, which is a minimal stochastic automaton describing the data stream. Therefore, no information on the system is required, except for the possibility of measuring the values of its relevant variables for a sufficiently long time interval. Crutchfield showed that the size of the $\epsilon$-machine diverges for systems at the phase transition, such as logistic maps at the onset of chaos~\cite{Str1994}.\\

The relation between evolution and criticality in CA has been first investigated by Packard~\cite{Pac1988} and subsequently inside the EvCA Project~\cite{MitCruHra1994,Hor2013}. An interesting finding of those studies is that the evolution of CA does not necessarily lead to the edge of chaos, as it might depend upon the evolutionary algorithm and the fitness function. This subject will be extensively discussed in Section~\ref{sec:discussion}.

The conjecture that critical systems achieve the highest level of information processing is supported by observing that in systems that undergo a phase transition information measures that are relevant for computation are maximized at the critical point. For example, Solé and Miramontes show that in an agent-based model in which agents move over a grid there exists a critical boundary in parameter space where maximum information transfer occurs~\cite{SolMir1995}. The study suggests that also natural systems composed of many interacting units---such as ant colonies---which have to coordinate so as to attain nontrivial goals, may have evolved towards critical dynamical regimes.

Subsequently, information-theoretic measures~\cite{CovTho2006} have been thoroughly applied with the aim of providing evidence for this hypothesis. The application of these methods makes it possible to quantitatively study information processing in complex systems and characterize the peculiarities of the dynamical regimes.
Some of these results concern Boolean networks (BNs), which will be often mentioned in this contribution for their relevance in this context. BNs were introduced by Kauffman~\cite{Kau1969,Kau1993} as a genetic regulatory network model and they have been shown to reproduce significant properties of complex systems. Some notable properties of BNs as models of genetic regulation will be surveyed in the following sections. Here, we briefly introduce them so as to summarise the basic notions for assessing the results on BNs computational properties that will be reviewed.
BNs are networks of binary automata, ruled by Boolean transition functions, which in general may be different for each automaton. Usually, an automaton in the network is called node. For a BN with $N$ nodes $x_1, x_2, \ldots, x_N$, the $N$-ple of node values $[x_1(t), x_2(t), \ldots, x_N(t)]$ at time $t$ represents the state of the network at time $t$. Nodes are supposed to update their state at discrete time steps. In the case of synchronous deterministic dynamics, there is only one successor for each network state, therefore the network starts from an initial state and evolves in time until it encounters a state already visited\footnote{Under the hypothesis that $N < \infty$} and then it repeats the same sequence of states. This sequence is called cycle or attractor and the cardinality of the states that compose it is the cycle length (or period). A special case is represented by cycles of length 1, usually called fixed points. The portion of the BN trajectory before the cycle (which can be empty) is called transient. The set of initial states that lead to a given attractor $\mathcal{A}$ is called the basin of attraction of $\mathcal{A}$.
The most prominent class of BNs is that of Random BNs (RBNs), in which functions and connections are chosen according to pre-defined distributions. A special case is the one in which nodes receive exactly $K$ distinct inputs chosen at random among the other nodes and each transition function is defined by composing the truth table assigning a 1 to each of the $2^K$ entries with probability $p$ (called bias). Nodes update their value in parallel and synchronously. RBNs show a phase transition between order and chaos depending on the values of $K$ and $p$~\cite{DerPom1986,ShmKau2004}. For $2p(1-p)K < 1$ RBNs have on average an ordered behaviour, whilst for $2p(1-p)K > 1$ the networks show extreme sensitivity to initial conditions and very long cyclic attractors, which denote a chaotic behaviour. For this BN model, the critical regime is achieved for $2p(1-p)K = 1$, which describes a curve in the $(K,p)$-plane, called the critical line. In addition, there exist also other ways to let BNs attain a critical regime, such as the choice of boolean functions with specific characteristics.\\

The transition between order and chaos in RBNs has been extensively studied by means of information-theoretic measures. R\"am\"o et al.~\cite{RamKauKesYli2006a} propose to use the Shannon entropy of the perturbation size distribution in RBNs as a measure for information propagation: this measure is maximized at the phase transition. Intuitively, this results supports the hypothesis that critical systems (critical RBNs in this case) have the largest repertoire of information propagation actions, without incurring in chaotic behaviours. Ribeiro et al.~\cite{RibKauLloSam2008} compute the average pairwise mutual information between nodes at subsequent time steps. The mutual information between two random variables measures the amount of information that the knowledge on one variable carries about the other. As a consequence, the mutual information between two nodes at subsequent time steps estimates the information transfer between nodes. Ribeiro et al. show that this measure is maximized along the critical line. Therefore, critical RBNs seem to attain a more efficient information transfer mechanism than that of ordered and chaotic RBNs. Krawitz and Shmulevich~\cite{KraShm2007} study the distribution of basins of attraction size in RBNs and find that the Shannon entropy of this distribution scales with system size only along the critical line, suggesting that the informationally optimal partition of the state space is indeed attained when the system is operating between order and chaos. As a consequence, only in critical RBNs size can scale with the capability of performing increasingly diverse and coordinated behaviour. Further evidence for the computational capabilities of critical RBNs is provided in~\cite{GalNykCarPri2010,MakKesKauYliNyk2011}, where set-based complexity is considered; this quantity measures the amount of significant information embedded in a set of elements. For example, let us take binary sequences of length N. A set of identical sequences carries negligible information (as redundancy is maximal) and the same holds for a set of completely random sequences, which carry no structure whatsoever. Galas et al.~\cite{GalNykCarPri2010} compute the set-based complexity of trajectories of RBNs in ordered, chaotic and critical regimes. They show that set-based complexity is maximized for critical RBNs. 

A principled approach for studying the computational capabilities in dynamical systems is provided by Lizier in his Ph.D. thesis~\cite{Liz2013}. The work by Lizier is particularly important because it makes it possible to quantitatively address some conjectures on the computational capabilities of complex systems. He studies information processing in terms of information storage, modification and transfer by using information-theoretic measures. 
A notable result concerns RBNs, for which information storage and transfer are studied across the dynamical regimes. Lizier and collaborators find that the dynamics of ordered RBNs is dominated by information storage, which increases moving towards the edge of chaos and then it decreases after the critical line. Information transfer also increases from order to disorder and peaks just inside the chaotic regime, thus disrupting the information storage capability. RBNs in the critical regime attain the optimal balance between these two capabilities~\cite{LizProZom2008}. As genetic regulatory mechanisms are often modelled by means of RBNs, these findings suggest that biological cells indeed evolved towards criticality so as to maximize coherent yet expressive computation.

To conclude this succinct survey on information processing in critical RBNs, we mention the fact that critical RBNs maximize Fisher information, which is known to be maximal in order parameters for systems at the phase transition~\cite{WanLizPro2011}.\\

The relation between critical regimes and computational capabilities has been studied also in systems other than CA and BNs. Kinouchi and Copelli~\cite{KinCop2006} propose a model of interacting neurons with random topology. A neuron can be in one out of three states: active, inactive and refractory. Each neuron can be activated either owing to an external stimulus or via the action of a neighbouring neuron active at the previous step, with probability $p_{ij}$. Neurons are connected randomly (forming an Erd\"os-R{\'e}ny graph) and connection weights $p_{ij}$ are random variables with uniform distribution, with $p_{ij}=p_{ji}$.\footnote{The symmetric coupling is chosen with the purpose of modelling electric gap junctions.} The overall activity of the system is measured as density of active nodes, i.e. the fraction of active neurons. Kinouchi and Copelli show that this model has a phase transition in the density of active nodes as a function of the average branching, which averages the weights of the network. For low average branching ratios, the extinction times of perturbations are low, whilst for high values of branching ratio the network indefinitely self-sustains the perturbation. Notably, critical networks have the largest variance in the distribution of extinction times and present a power law behaviour in the distribution of avalanche sizes.
Bertschinger and Natschl\"ager~\cite{BerNat2004} study networks of randomly connected threshold gates and show that they exhibit a transition between ordered and disordered dynamics depending on the connectivity of the network. The networks found at the edge of chaos are those able to perform complex computations on time series. A similar result is presented in~\cite{LegMaa2007}, where networks of spiking neurons are studied. 

The findings previously surveyed support the hypothesis that reliable and flexible computational capabilities are a general property of critical dynamical systems. Further evidence comes from the field of optimization, where some results suggest that also the best problem solving capabilities are attained at a phase transition, at the edge of chaos. In particular, it has been shown that the performance of local search algorithms is maximized when a parameter controlling the parallelism of local moves is properly tuned~\cite{MacSiaKau1996,KauMac1995,Rol2001}. Macready et al.~\cite{MacSiaKau1996} have shown that this value is indeed critical and corresponds to a phase transition in the entropy of the system.

%%%%%%%%%%%%%%%%%%%%%%%%%%%%%%%%%%%%%%%%%%%%%%%%%%%%%%%%%%%%%%%%%%%%%%%%%%%%%%%%%%%%%%%%%%%%%%%%%%%%%%%%%%%%%%%%%%%%%%%
\section{Critical living systems}
\label{sec:criticalLivingSystems}

It has been conjectured that systems in critical regime have advantages over systems totally ordered or disordered and that this condition is achieved during evolution. Inspiring discussions on this subject can be found in works by Kauffman~\cite{Kau1993,Kau1996}, where this tantalizing hypothesis is proposed and discussed, along with preliminary yet significant results. According to Kauffman, systems at the edge of chaos attain the best balance between robustness and adaptiveness; furthermore, they are able to ``coordinate past discriminations with reliable future actions'' (quoted from~\cite{Kau2008}), i.e. they reliably and robustly respond to inputs while being able to react with a wide repertoire of possible actions.
In the last ten years, compelling results have attracted much interest and revitalized research on the subject.
Some researchers have addressed the question as to whether cells are critical, achieving notable results~\cite{SerVilSem2004,SerVilGraKau2007,RolVerSer2010,ShmKauAld2005,RamYli2006b,NykPriAldRamKauHooYliShm2008a,Balleza2008,ChoLloSmoBaiHugYliChuRib2010}  by comparing statistical properties of ensembles of genetic regulatory network models with statistical properties of real cells. The rationale of these approaches relies in the comparison of statistical properties of ensembles of biological genetic networks and RBNs or similar models: the best fit is attained when the models are drawn from a distribution in which the parameters assume the critical value, i.e. the one that separates the ordered from the disordered phase. This is indeed the method used to identify dynamical criticality: a model of the system is built and its dynamical regimes are studied as a function of one or more control parameters. Data from ordered and disordered regimes are collected, as well as data at the border between the two regimes. The data produced by the model are then compared: if the best match is achieved when model parameters have the critical value, then evidence for the criticality of the real system is found.
In particular, the work by Serra and collaborators~\cite{SerVilSem2004,SerVilGraKau2007} deals with avalanches in the expression of genes produced by gene knock-out. They show that critical RBNs are the ones achieving the best fit with real data from microarray experiments on the \textit{S.~Cerevisiae}. A crucial role in that work is played by the so-called Derrida parameter $\zeta$, which is an index of the dynamical regime of discrete systems like BNs~\cite{Kau1996}. For  $\zeta < 1$ a single node perturbation affects in one step less than one other node on average (ordered regime), whilst for $\zeta > 1$ the perturbation reaches more than one other node in one step (disordered regime); the condition $\zeta = 1$ identifies the critical regime. In~\cite{SerVilGraKau2007,DiStefano-wivace2015} it is shown that the distribution of avalanches depends only on the Derrida parameter and that the best match between data from \textit{S.~Cerevisiae} and RBNs is achieved for a value of $\zeta$ slightly less than 1; other values of $\zeta$ would lead to quite different avalanches distributions.
This result is in agreement with the ones attained by Shmulevich et al.~\cite{ShmKauAld2005} who compare the Lempel-Ziv complexity~\cite{LemZiv1976} of data stream generated by a genetic regulatory network model of the HeLa cells with that of RBNs in different regimes. They find that critical or slightly sub-critical RBNs are those with the best correspondence with biological data. A similar approach has been followed by Nykter et al.~\cite{NykPriAldRamKauHooYliShm2008a}, who compared time series of macrophages and of RBNs by means of the normalized compression distance~\cite{LiCheLiMaVit2004} and show that the best fit is attained with critical RBNs.\\
 
Balleza et al.~\cite{Balleza2008} show that genetic regulatory network models of several organisms are critical, to the extent that, when perturbed, their behaviour is the same of critical RBNs.
Chowdhury and collaborators~\cite{ChoLloSmoBaiHugYliChuRib2010} infer a BN model describing the S. Cerevisiae and find that the resulting BN has some characteristics of critical networks. Analogous results are shown by Darabos et al.~\cite{DarGiaTomProDi2011}, while Hanel et al.~\cite{HanPocThu2010} show that even a simple genetic regulatory model containing a minimal nonlinear contribution can be tuned at the edge of chaos, suggesting that many models can indeed enjoy the same property.
Finally, indications about the presence of critical values in natural cells are also provided in~\cite{Kan2006}.  

These results are not conclusive, but they anyway support the hypothesis that biological cells are in a dynamical regime between order and chaos. 

Besides cells, also other biological systems have been studied and the results suggest that they may enjoy the same property. A striking example is that of neural dynamics: notable results and models have been proposed in neuroscience, such as the ones discussed in~\cite{Beg2008,Chi2010,FriItoet2012,TagChi2013}. These results bring evidence to the hypothesis that brain dynamics is critical, as they show power law in the distribution of avalanches only when neurons are at the normal activity condition, whilst they behave differently when they are kept over or under activated. As stressed in~\cite{BegTim2012}, these findings suggest that the brain is indeed dynamically critical, i.e. it is poised between an ordered and a disorder regime.

Critical dynamics has also been found in models of flocks of birds~\cite{MorBia2011} and in morphogenetic processes~\cite{KroDub2014}.\\

It is worth to mention that the notion of ``extended critical situations'' has been proposed to describe the case of living systems~\cite{BaiLon2008}. The intuition of this proposal is that biological entities permanently keep themselves in a region of criticality, rather than a point.\\

Experiments on real biological systems and results on models provide strong evidence to support the ``criticality hypothesis''. Yet, the reasons why biological systems seem to be poised between order and disorder are still unclear and somehow elusive. Indeed, there are two main noncontradictory views of this conjecture:

\vspace{1ex}
\textit{(i)} The first one can be stated saying that critical systems are more evolvable than systems in other dynamical conditions because they optimally balance mutational robustness (i.e. mutations just slightly change the phenotype, without introducing dramatic changes) and phenotypic innovation (i.e. mutations can introduce significant novelty in the phenotypes).

\vspace{1ex}
\textit{(ii)} The second version of this conjecture refers, instead, to the possible fitness advantages of critical systems over ordered or disordered ones, which seem to be the ability of attaining the most effective balance among information storage, modification and transfer, and of achieving the best trade-off between the repertoire of their possible behaviours and reliability of their actions.

\vspace{1ex}
Several attempts have been made for providing support to the conjecture, in either version. All these works share a similar approach, which consists in studying the artificial evolution of some model, such as Boolean networks. We remark that in this survey we are interested in those works directly addressing the criticality hypothesis, rather than the evolution of notable properties such as robustness, adaptivity, modularity or specific topological features.

Aldana et al.~\cite{AldBalKauRes2007} address the problem of the relation between robustness and evolvability and provide evidence to support that critical BNs achieve an optimal balance between these two properties. The work by Torres-Sosa et al.~\cite{TorHuaAld2012} strengthens this result by analysing the outcome of an artificial evolutionary process. In their work, RBNs are subject to artificial Darwinian evolution operating by means of mutation and gene duplication. Selection favours networks that are able to both \textit{(a)} maintain the current ``phenotypes'' (i.e. attractors) and \textit{(b)} generate new ones. The authors show that this evolutionary process drives RBNs towards the edge of chaos.
Further evidence to the evolvability of critical RBNs is provided in~\cite{NykPriLar2008b} where RBNs at the edge of chaos are shown to be maximally diversified in their structure. Similar results concerning the diversity among attractors in critical RBNs have been presented in~\cite{RolSer2011b}.

A key factor in the relation between robustness and evolvability is the fitness landscape on which the evolutionary process acts. In a seminal work, Kauffman and Smith~\cite{KauSmi1986} investigated the relation between the parameters of RBNs (connectivity and bias) and the properties of the evolution landscape. This link has been investigated explicitly w.r.t. the criticality hypothesis in~\cite{BenVilRolSerManGagPinBir2013}, where it has been shown that not all the tasks for which a BN is trained necessarily lead to a critical regime. In fact, on the one hand the dynamics of a RBN influences its mutational robustness and its phenotypic plasticity, but on the other hand it may happen that the properties of the fitness landscape are dominated by the kind of task for which networks are selected. This is indeed a crucial point in the criticality hypothesis, which has not yet sufficiently investigated and that will be discussed in Section~\ref{sec:discussion}.
The interplay between selective pressure and dynamical regime of RBNs has also been investigated in a co-evolutionary settings in~\cite{Kau1991}, where BNs are evolved to play the mismatch game, i.e. a two-players game in which networks should compare some preselected nodes and try to match the negated values of the opponent. It was shown that evolution produces BNs that are critical. 
In~\cite{KauJoh1991} an abstract model of species co-evolution is studied, in which the landscape of one species changes the landscape of the other during the evolutionary process. In particular, the NK-model is used~\cite{Kau1996}. Results suggest that evolution leads to an equilibrium in which evolutionary avalanches appear to propagate on all length scales in a power law distribution.
Christensen et al.~\cite{ChrDonKoiSne1998} study the evolution of random networks of interacting elements (a model similar to the NK-model) under extremal dynamics~\cite{BakSne1993} and find that the evolutionary process lead to networks with critical connectivity. Analogous results are presented in~\cite{BorRoh2000} and~\cite{LiuBas2006} for threshold networks and co-evolving RBNs, respectively.\\

The second stream of studies concerns the investigation of the evolutionary advantages of critical systems owing to their enhanced computational capabilities. In Section~\ref{sec:edgeOfChaos} we have provided an overview of the results suggesting that optimal computational capabilities are attained at the edge of chaos. Nevertheless, it is also important to test these hypotheses in evolutionary contexts. 
Goudarzi et al.~\cite{GouTeuGulRoh2012} observe that RBNs evolved to be able to solve combinatorial tasks of varying hardness (e.g. binary addition and even-odd classification) converge to populations of critical networks. Indeed, such BNs exhibit a damage spreading behaviour typical of critical RBNs. Interestingly, the authors find that for all the three tasks considered in their work, evolution leads to criticality. The discrepancy between that work and~\cite{BenVilRolSerManGagPinBir2013} suggests that the choice of the fitness function may play a major role in the outcome of the experiments and provides indications for further investigations.
Hidalgo et al.~\cite{HidGriSuwMunBanMar2014} study the evolution of agents whose behaviour is modelled by means of probability distributions over their possible actions. The fittest agents are those which best react to the external stimuli. For simplicity, it is supposed that both environmental stimuli and agent actions are represented as binary strings: the best action for an agent is a string completely matching the environmental string. Agents have internal parameters controlling their actions distribution, which should match the stimuli distribution of the environment (of course, it is supposed that a perfect match is not possible, as e.g. the number of agent internal parameters is lower than that of environmental parameters). Three scenarios are studied: \textit{(1)} agents face a static environment; \textit{(2)} agents face a variable environment, composed of many different stimuli sources; \textit{(3)} agents interacts among themselves, so the environment for an agent is composed of the other agents. The evolutionary process acts on the parameters of the agent action probability distribution. Results suggest that the critical regime is a stable evolutionary solution when agents try to optimize their interaction with a changing environment or among themselves, whilst for low complexity environments the systems tend to remain non-critical. This result reinforces the hypothesis that criticality is an evolutionary advantage only under some conditions, which may involve variability, dynamicity and complexity.

%%%%%%%%%%%%%%%%%%%%%%%%%%%%%%%%%%%%%%%%%%%%%%%%%%%%%%%%%%%%%%%%%%%%%%%%%%%%%%%%%%%%%%%%%%%%%%%%%%%%%%%%%%%%%%%%%%%%%%%
\section{Discussion}
\label{sec:discussion}

Despite the promising results achieved so far on the criticality hypothesis, some important issues and open questions have still to be addressed.\\

First of all, different definitions of critical system have been used in the literature. For example, most definitions of criticality rely on the properties of the effect of perturbations on the systems, but the condition of the systems when perturbed and how to measure the effects of perturbations are often implicit~\cite{CamVilPolSer2011,VilCaDamRolFilSer-natural-computing}. This makes the ground of the discussions slippery and might induce unsound conclusions.
In order to test the criticality hypothesis, it is necessary to state it in more precise terms. In general, criticality is related to the average behaviour of a small perturbation of a system state. In the ordered case, two systems with the same parameters but slightly different initial conditions will, on average, tend to the same state, so that their differences will die out, while in the disordered (chaotic) case the distance between the two systems will initially tend to grow. Critical systems are intermediate between these two cases, so the size of the perturbation remains initially constant on average.
Note however that when we consider the ``average'' behaviour of a perturbation we may actually refer to different types of ensembles over which the averages are taken. Failing to appreciate these differences may lead to misunderstandings and erroneous conclusions, as it has been demonstrated~\cite{CamVilPolSer2011,VilCaDamRolFilSer-natural-computing}.
Averages can be taken for example on a particular instance of the system under examination (i.e. keeping fixed the form of equations and the values of the parameters); in this case the average can be taken either over all the possible initial conditions or over the different initial conditions that lead to the same attractor (i.e. a state or a set of states that are reached from some initial condition after the transients have died out). However, it has been shown that averages taken over completely arbitrary initial states can be misleading, as some of these states might be impossible to reach under any conditions. It is therefore interesting also to consider a restricted set of initial conditions, for example limiting to those that can be the successor of another (arbitrary) state.
Moreover, it is also interesting to consider averages taken only over the states that belong to a specific attractor, or over the states that belong to an attractor (whichever it is).
Last but not least, it is important to consider also ensembles of systems with different parameter values. We will refer to this kind of averages as to the structure ensemble averages, that can be taken according to the same different alternatives that have been described above for the case of a single instance of the system.
As it has been observed, criticality is related to the average behaviour of a small perturbation, and the different ensembles induce different (but related) definitions of criticality. This notion has been quantified in that of sensitivity for discrete dynamical systems and BNs in particular~\cite{ShmKau2004}. The sensitivity of a Boolean function measures how sensitive the output is with respect to changes in the inputs. Sensitivities can be computed on the various ensembles mentioned above, and the size of the ensembles over which averages are taken naturally leads to a hierarchy of sensitivities.\\

The identification of the conditions under which evolutionary dynamics favours critical systems is still a fundamental open question. As previously mentioned, recent works have addressed this point~\cite{BenVilRolSerManGagPinBir2013,HidGriSuwMunBanMar2014}, but this issue still requires a thorough and principled investigation. In particular, this requires the definition of a function that measures the effectiveness with which a system accomplishes a given task, i.e. a fitness function. Generally speaking, in the case where dynamics is dominated by attractors, the different attractors will define, for a given system and set of parameter values, its dynamic repertoire, i.e. the different states that it can reach starting from various conditions. Therefore the fitness function will often be related to the attractor landscape. We expect, on the basis of reasonable guesses and of previous studies, that critical systems will not always be preferred, since the outcome depends upon task’s features (see~\cite{BenVilRolSerManGagPinBir2013}). Therefore the goal is indeed quite complicated, as it amounts to finding out for which kind of tasks critical systems are better. Our guess is that this is more likely to happen in time-varying complicated tasks. The different types of tasks should therefore be categorized in a proper way, in particular by the value of some parameters that describe the main characteristics of the fitness landscape (e.g. correlation lengths) and of its own dynamics (e.g. the rate of change and its amount).

The last observation is tightly related to the connection that the system has with its environment and the way by which it interacts with it. Indeed, the importance of the environment on system criticality and of the openness of the systems are often overlooked. In addition, the relation among evolution, adaptation and learning is often just informally addressed. All these issues are in fact relevant for the identification of the conditions under which systems evolve towards critical regimes and to understand the reasons of this phenomenon.\\

A further issue that has not yet been discussed above concerns artificial systems: if critical dynamical systems have evolutionary advantages in nature, then this property may also hold for artificial systems, such as learning ones. Therefore, enforcing dynamical criticality or correlated properties may provide a general criterion for the automatic design of such systems. This would complement the usual approach consisting in defining ad-hoc and task dependent fitness functions in evolutionary techniques. Some preliminary work has been done aimed at investigating the relation between fitness and some information-theoretic measures~\cite{SpeTriNol2008,EdlChaHinKocTonAda2011,JosTonKoc2013,RolBenBirPinSerVil2011a}, but without explicitly considering dynamical criticality. On the one hand, the exploration of this idea may pave the way for devising advanced methods both for learning techniques and system design, pushing the envelope on the design of autonomous open systems. On the other hand, it would contribute to deepen our knowledge on the criticality hypothesis itself.\\

Last but not least, it is important to observe that the question as to whether critical systems share some common properties in their internal organization has not yet been addressed. It has been shown that information-theoretic measures make it possible to detect dynamical structures in complex systems~\cite{VilFiBenRolLanSer2013-ALIFE,filisetti-ecal2015,Villani-wivace2015}. The method is based on a measure called the dynamical cluster index and can detect subsets of variables that are tightly integrated among themselves and loosely interacting with the rest of the systems. Therefore, this method may provide an effective tool for identifying common characteristics in the organization of critical dynamical systems, in which the structure and hierarchy of relevant subsets are expected to be different from those in ordered and disordered regimes.

%%%%%%%%%%%%%%%%%%%%%%%%%%%%%%%%%%%%%%%%%%%%%%%%%%%%%%%%%%%%%%%%%%%%%%%%%%%%%%%%%%%%%%%%%%%%%%%%%%%%%%%%%%%%%%%%%%%%%%%
\section{Conclusion}
\label{sec:conclusion}

In this paper we have provided an overview of dynamical criticality, as it is discussed in the natural sciences and computer science. Evidence of a dynamics between order and disorder has been found for systems such as biological cells and the brain; moreover, notable results support the conjecture also computational systems across the critical regime are capable of attaining an optimal trade-off between reliability and flexibility. We have also briefly outlined some open questions on dynamical criticality that have still to be addresses, concerning foundational aspects and possible applications in artificial systems design.

%%%%%%%%%%%%%%%%%%%%%%%%%%%%%%%%%%%%%%%%%%%%%%%%%%%%%%%%%%%%%%%%%%%%%%%%%%%%%%%%%%%%%%%%%%%%%%%%%%%%%%%%%%%%%%%%%%%%%%
\section*{Acknowledgements}
We are deeply indebted to Stuart Kauffman for his inspiring ideas and for several discussions on various aspects of the criticality hypothesis. We also gratefully acknowledge useful discussions with David Lane, Alex Graudenzi and Chiara Damiani.

%%%%%%%%%%%%%%%%%%%%%%%%%%%%%%%%%%%%%%%%%%%%%%%%%%%%%%%%%%%%%%%%%%%%%%%%%%%%%%%%%%%%%%%%%%%%%%%%%%%%%%%%%%%%%%%%%%%%%%

\end{document}